\documentstyle[psfig]{aipproc}

\newcommand{\vol}[2]{$\;\,$\bf #1\rm , #2.} 
             \font\sevenrm=cmr7

               \font\sixbf=cmbx6      
\def\aap{{\it Astron. Astr.}}                         
\def\aa{{\it Astron. Astr.}}

\def\apss{\it Astrophys. Sp. Sci.}
\def\djpG{{\it J. Phys. G: Nucl. Part. Phys.}}
\def\phrep{\it Phys. Reports}
\def\pasp{\it Pub. Astron. Soc. Pacific}
\def\ssr{{\it Space Sci. Rev.}}                       
\def\apj{{\it Astrophys. J.}}                         
\def\apjl{{\it Astrophys. J. (Lett.)}}                
\def\apjs{{\it Astrophys. J. Supp.}}                  
\def\mnras{{\it MNRAS}}                               
\def\nat{{\it Nature}}                                

\def\laeq{\;\raise.2ex\hbox{$<$}\kern-.75em\lower.9ex\hbox{$\sim$}\;}
\def\gaeq{\;\raise.2ex\hbox{$>$}\kern-.75em\lower.9ex\hbox{$\sim$}\;}
\def\teq#1{$\, #1\,$}                           
{\catcode`\@=11                                                  
\gdef\SchlangeUnter#1#2{\lower2pt\vbox{\baselineskip 0pt\lineskip0pt    
\ialign{$\m@th#1\hfil##\hfil$\crcr#2\crcr\sim\crcr}}}}           
\def\gtrsim{\mathrel{\mathpalette\SchlangeUnter>}}

\def\pmb#1{\setbox0=\hbox{#1}%
  \kern-0.0125em\copy0\kern-\wd0
  \kern0.025em\copy0\kern-\wd0
  \kern-0.0125em\raise0.0433em\box0 }

\def\Emax{E_{\hbox{\fiverm MAX}}}

\def\Emax{E_{\hbox{\sevenrm max}}}
\def\jour#1#2#3#4{#4, {#1} {\bf #2}, #3}

\begin{document}


\title{Supernova Remnants and Plerions\\
      in the Compton Gamma-Ray Observatory Era}

\author{Ocker C. de Jager$^*$ and 
        Matthew G. Baring,$^\dag$\thanks{Compton Fellow, Universities Space
        Research Association} }
\address{$^*$Space Research Unit, PU vir CHO, Potchefstroom 2520,
         South Africa\\
         $^\dag$Laboratory for High Energy Astrophysics, Code 661,\\
         NASA Goddard Space Flight Center, Greenbelt, MD 20771, USA}
\maketitle

\begin{abstract}
Due to observations made by the Compton Gamma-Ray Observatory over
the last six years, it appears that a number of galactic supernova
remnants may be candidates for sources of cosmic gamma-rays.  These
include shell-type remnants such as IC443 and $\gamma$ Cygni, which
have no known parent pulsars, but have significant associations with
unidentified EGRET sources, and others that appear to be composite,
where a pulsar is embedded in a shell (e.g. W44 and Vela), or are
purely pulsar-driven, such as the Crab Nebula.  This review discusses
our present understanding of gamma-ray production in plerionic and
non-plerionic supernova remnants, and explores the relationship between
such emission and that in other wavebands.  Focuses include
models of the Crab and Vela nebulae, the composite nature of W44, the
relationship of shell-type remnants to cosmic ray production, the
relative importance of shock-accelerated protons and electrons,
constraints on models placed by TeV, X-ray and radio observations, and
the role of electrons injected directly into the remnants by parent
pulsars.  It appears as if {\it relic electrons} may be very
important in the Vela and Crab remnants.  The recent observation of the
TeV hot spot in the Vela remnant, which is offset from the current
pulsar position, is attributed to relic electrons that were left at the
birthplace of the pulsar, the offset being due to the proper motion of
the Vela pulsar during its 11,000 year lifetime.  We also discuss the
role of {\it freshly-injected} electrons in the remnants around the
Crab, Vela and PSR B1706-44 pulsars.  These electrons can acquire
energies that tap up to at least 10 percent of the full pulsar polar
potential, and can produce prominent synchrotron and inverse Compton
radiation signatures.  The various recent models for predicting
gamma-ray emission in shell-type remnants are summarized.  The
constraining upper limits to TeV emission from such remnants obtained
by the Whipple Observatory indicate that either the emission due to
particles accelerated at remnant shocks is too faint to be detected by
EGRET, or that conditions near their shells (e.g. high density, low
magnetic field) limit the acceleration of particles to below a few TeV.
\end{abstract}

\section*{Pulsar-Driven Nebulae (Plerions).}

A number of plerions have been discovered in radio, optical, and X-rays
\cite{seward89}, with the Crab as the youngest and most energetic
source.  Plerionic nature is usually indicated by a center-filled
morphology, resulting from the continuous injection of pulsar electrons
into the nebula, with the additional constraint that the spectra must
be non-thermal (power-law) resulting from a statistical acceleration
processes.  The reason for the latter constraint is that some
center-filled X-ray remnants show evidence for thermal emission such as
W44 \cite{jsa93}, resulting from the presence of hot gas from the
central regions.

The radio, optical and X-ray emission observed in plerions is believed
to be due to synchrotron emission.  Observations map the product of the
field strength (to some power) and the number of energetic particles
via the synchrotron brightness, however the spectral components from
different wavebands cannot be separated unless another emission
mechanism is observed from the same particle population.  An example of
this is inverse Compton scattering, where the magnetic field is
replaced with a photon density as a ``target'' for the high energy
particles, and observations of both processes lead to a determination
of the field strength.  Inverse Compton emission is usually seen at
higher energies compared to synchrotron emission, and a multiwavelength
study of a source allows us to simultaneously probe the synchrotron and
inverse Compton processes, thereby permitting measurement of both the
magnetic field strength and the total energy budget of the electrons.
In fact, Compton Gamma-Ray Observatory (CGRO) observations of the Crab
Nebula provided the first detection of the transition from synchrotron
to inverse Compton emission \cite{dea96a}, whereas CGRO observations of
the Vela remnant are limited to the OSSE instrument (see
Fig.~\ref{fig:Vela}), and therefore allow us to probe only the
synchrotron component of the Vela spectrum.

\subsection*{Constraints on the development of a plerion} 

The first condition for the development of a synchrotron nebula around
a pulsar is that some fraction of the Poynting flux from the pulsar
should be transferred to electrons. This means that the ratio $\sigma$
of Poynting to particle energy fluxes should not be too large.  The
second condition is that either a shock (at an angular distance $r_s$),
or an instability in the wind (at an angular distance $r_o$) should
remove the electrons from the comoving frame of the pulsar wind, so
that the electrons can ``see'' an effective perpendicular field
component for synchrotron radiation \cite{kc84}.  For example, high
resolution HST images of the Crab pulsar/nebula shows that the first
knots (disturbances in the relativistic unperturbed flow) are seen at a
distance of $r_o<1"$ on its polar axis \cite{hester95}, whereas the
pulsar wind shock is expected at $r_s\sim 8"-10"$  \cite{kc84}.  The
combination of these two effects (energy transfer and the
``visibility'' of a field for synchrotron radiation) results in
plerionic synchrotron emission outside a minimum distance from the
pulsar. The development of a shock also allows the pulsar wind to be
slowed down from relativistic speeds to the speeds of typical expanding
supernova ejecta.

\subsection*{The Crab Nebula}

The Crab Nebula is the prototype cosmic source of synchrotron radiation
and inverse Compton (IC) scattering \cite{gould65,weekes89,dh92}. Since
synchrotron photons are inverse Compton scattered by their parent
relativistic electrons, the Crab Nebula is considered to be a
synchrotron-self-Compton source.

\subsubsection*{The Structure of the Nebula}

ROSAT and HST observations of synchrotron emission from the Crab Nebula
led Hester et al.\cite{hester95} to make the following fundamental
observation about the structure of this nebula: almost all observations
of the system at all scale sizes show a well-defined axis of
cylindrical symmetry running from the southeast (SE) to the northwest
(NW) through the center of the nebula, at an angle tilted by
$20^{\circ}-30^{\circ}$ with respect to the plane of the sky. This axis
corresponds to the direction of elongation of the nebula as a whole,
the axis of the X-ray and optical jets, the X-ray torus (first
identified by \cite{ab75}) and the alignment of the optical ``wisps.''

Hester et al.\cite{hester95} also summarized the main properties of
the pulsar wind which is responsible for the unpulsed emission over
several decades in energy.  The symmetry axis is probably associated
with the pulsar spin axis, and the DC component of the rotating
magnetic field results in a helical polar wind centered on the spin
axis. The elongated optical synchrotron nebula appears to be associated
with this high latitude structure.  Equipartition between particle and
field energy is probably reached in the optical nebula, with $B_{\rm
optical}=3\times 10^{-4}$ G \cite{hester95}. This optical nebula is
also expected to be the source of IC TeV $\gamma$-rays
\cite{gould65,dh92}.  In fact, TeV observations of the Crab Nebula did
confirm a field strength of $\sim 3\times 10^{-4}$ G for the
optical/TeV nebula \cite{dh92}. Closer to the pulsar, \cite{hester95}
identified relatively small optical knots in the polar axis with
equipartition field strengths as high as 2 mG, and \cite{dea96a}
discussed the possibility that the variable structures near the pulsar
may be associated with the variable $\gamma$-ray emission seen by EGRET
(see also below).

The equatorial zone (identified as the X-ray torus by \cite{ab75})
extends $\sim \pm 10^{\circ}$ from the spin equator of the system, with
a relatively low field (azimuthally wound up) strength distribution of
$B_{\rm torus}\sim 7\times 10^{-5}(r/8")^{0.5}$ G for $r>8"$, as
derived by \cite{bork89} from EINSTEIN HRI images.  The X-ray emission
up to at least 50 keV is associated with this torus \cite{pelling87},
but above this energy we have no imaging capabilities to identify the
site of gamma-ray emission, and we have to rely on spectral and
temporal characteristics to infer constraints on the properties of the
gamma-ray emission site (see \cite{dea96a} for a more detailed
discussion of the gamma-ray properties).

\subsubsection*{The high energy synchrotron tail}

CGRO probed the energy range above 50 keV where no imaging is possible,
and a comparison between BATSE, OSSE, COMPTEL and EGRET observations of
the Crab total emission was made by \cite{much96}.  Whereas BATSE
overestimated the the low energy gamma-ray flux, OSSE and COMPTEL
produced consistent results below 1 MeV. The spectral steepening above
$\sim 100$ to 200 keV reported by \cite{jung89} and \cite{bartlett94}
is confirmed by OSSE observations. This indicates that the toroidal
component terminates gradually above $\sim 200$ keV.

Conflicting results are however produced above $\sim 1$ MeV: whereas
COMPTEL observed a consistent spectral hardening above 1 MeV, OSSE
observed this hardening only during Observation 221 \cite{much96}.  The
flux in the 1--30 MeV range also appears to be variable with time
\cite{much95}, and if this hardening is real, it would be indicative of
the presence of another $\gamma$-ray emitting site. The rapid
variability associated with this component was interpreted by
\cite{dea96a} as an emission site near the pulsar where $B\sim 0.1
\mu$G. Hester et al.\cite{hester95} identified the optical knots near
the pulsar as shock-like sites with $B$ of the same order as required
by De Jager et al. \cite{dea96a}.

A comprehensive analysis of EGRET observations of the Crab Nebula was
reported by \cite{dea96a} and it was shown that the hard COMPTEL
component should cut off above $\sim 25$ MeV to meet the steep EGRET
spectrum between 70 MeV and 150 MeV.  De Jager et al. \cite{dea96a}
interpreted the steep spectrum between 30 MeV and  150 MeV as the
exponential tail of the synchrotron cutoff in the Crab Nebula. The
e-folding energy at 25 MeV is consistent with the interpretation that
electron acceleration in a relativistic shock occurs at a rate equal to
the gyrofrequency. This acceleration is constrained by synchrotron
losses, and it was shown that the synchrotron characteristic energy
associated with the highest electron energy is independent of the
magnetic field strength, and depends only on fundamental constants and
a factor $\epsilon\sim 1$ which depends on the Doppler factor and the
average electron pitch angle. Thus,
\begin{equation}
h\nu_{\rm max}=\epsilon \Bigl({{3}\over{4\pi}}\Bigr)^2\, {{hc}\over{r_e}}
=\epsilon\, {{9}\over{8\pi}} \; {{m_ec^2}\over{\alpha_{\hbox{\sevenrm f}}}} 
\approx 25 \epsilon\;{\rm MeV},
\end{equation}
where $r_e$ is the classical electron radius and
$\alpha_{\hbox{\sevenrm f}}$ is the fine structure constant.
 
Furthermore, this EGRET component was also found to be variable on a
timescale similar to the COMPTEL variability timescale.  Thus, whereas
the stable emission below 1 MeV is known to be associated with the
torus, the variable hard component  associated with the synchrotron
cutoff leads to the conclusion of an association with a high-B region
close to the pulsar, which is removed from the torus. The polar region
where high-B optical knots and variable ``anvil'' is found, is therefore
a candidate region for this variable hard component.  However, future
$\gamma$-ray observations above 1 MeV with improved calibration is
required to confirm these findings.

%
\begin{figure}
\vspace{-0.1truecm}
\centerline{ \hskip 3.0truecm\psfig{file=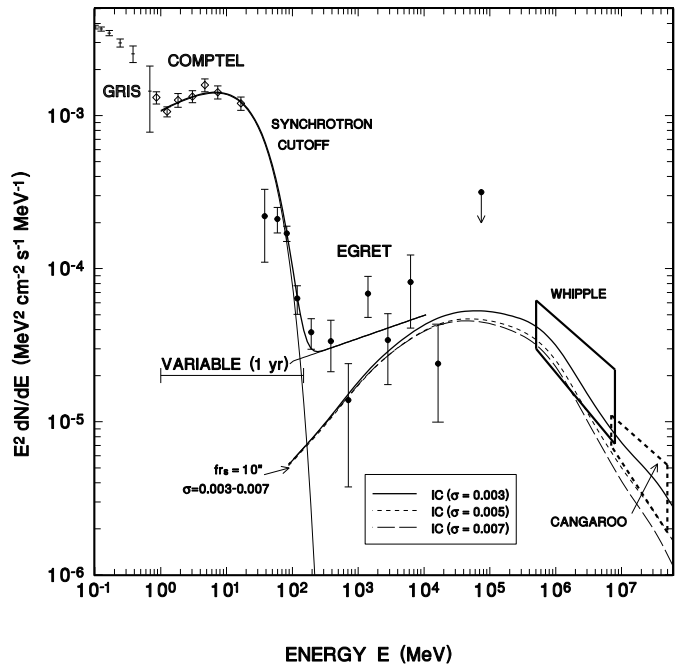,height=3.6in} }
\vspace{-0.0truecm}
\caption{The Crab nebular unpulsed $\gamma$-ray spectrum ($E^2dN/dE$) 
in the energy range 0.1 MeV to 20 TeV. The references are: 
GRIS [7],    
COMPTEL [61],    
\& EGRET [19].   
References to TeV points are given by [19].    
The Whipple error box (including systematic and statistical errors)
is from [16], 
and the CANGAROO error box is from [70].  
A two-component fit (1-150 MeV) resulted in a power law with
an exponential cutoff at 25 MeV, and an inflection point at 150 MeV. The 
inverse Compton model of [18] 
was used to generate spectra for
$r_s=10"$ and three values of $\sigma$ (as labelled),
as shown for energies between 100 MeV and 20 TeV. The model uncertainty is
a factor 2 at any energy. 
}
\label{fig:Crab}
\end{figure}

\subsubsection*{The Inverse Compton component}
Above $\sim 150$ MeV, \cite{dea96a} found a steady hard component,
which steepens gradually to meet the very high energy spectral points
above 100 GeV (see Fig.~\ref{fig:Crab}).  This steady hard component
was predicted to be the inverse Compton component. The {\it relic}
radio emitting electrons also scatter the optical background photons
into the $\sim 100$ MeV to GeV energy range, whereas the younger
optical and X-ray emitting electrons are expected to be responsible for
the TeV $\gamma$-rays. The lifetimes of the radio to X-ray emitting
electrons are much longer than the timescale of gamma-ray observations,
which implies that we should not expect to see a variable inverse
Compton $\gamma$-ray component.  The observed $\gamma$-ray flux above
150 MeV was found to be larger than predicted by the
synchrotron-self-Compton model \cite{dea96a}.  The EGRET $\gamma$-rays
above 150 MeV may be associated with inverse Compton $\gamma$-ray
emission if the field strength in the larger radio nebula is $\sim
1.3\times 10^{-4}$ G, which is smaller than the field strength in the
smaller optical/TeV nebula. This would suggest a departure from the
nebular field distribution derived by \cite{kc84}.

\subsection*{The Vela Supernova remnant}
The Vela Supernova remnant appears to be associated with the Vela pulsar
PSR B0833-45, which has a spindown age of about 11,000 years.
A large-scale {\it ROSAT} image of thermal soft X-rays ($kT=0.12$
keV) from the Vela SNR shows that the diameter of the remnant is
$7.3^{\circ}$, with the pulsar at its center.
In hard X-rays we do not see this thermal shell. Rather, the synchrotron 
nebula is resolved into an elongated (NE-SW) hard X-ray (2.5 - 10 keV)
structure as shown by \cite{willmore92}, and a compact nebula 
surrounding the pulsar. The energy-dependent geometry of this remnant
was illustrated in Fig. 1 of \cite{dhs96}.

\subsubsection*{The OSSE detection of Vela}
Fig.~\ref{fig:Vela} shows the non-thermal X-ray to low energy gamma-ray
spectrum of the 1 arcmin compact synchrotron nebula around the Vela
pulsar as seen by ROSAT, EINSTEIN, Birmingham and possibly OSSE as
reported by \cite{dhs96} and \cite{strickman96}.  Even though the OSSE
instrument does not have imaging capabilities for a clear association,
the spectra of other X-ray sources in the field-of-view cut off below
the OSSE range, and only the Vela SNR remains as candidate.
Furthermore, since the OSSE spectrum connects smoothly with the lower
energy X-ray (imaging) spectra, the association is quite likely.  The
authors have also shown that the energetics of electrons (given the
field strength scaled from the pulsar) is sufficient to produce
$\gamma$-rays into the low energy part of the EGRET range
(Fig.~\ref{fig:Vela}).  The unbroken spectrum and the EGRET upper limit
constrain the maximum electron energy to values smaller than those
expected from the $3\times 10^{15}$ volt polar cap potential drop.
More sensitive observations above 400 keV are needed to measure the
cutoff energy, which should lead to a measurement of the maximum
electron energy in the nebula.

\subsubsection*{An inverse Compton component from Vela X?}
The relatively hard spectrum associated with the bright Vela X radio
nebula (about 1$^{\circ}$ south of the pulsar) represents synchrotron
emission from relic electrons in the nebula, while the inverse Compton
emission resulting from the scattering of the 2.7K cosmic microwave
background produces MeV to GeV $\gamma$-rays.  This component would be
observable if the electron concentration in the Vela X remnant were
sufficient. However, we only know the synchrotron brightness of Vela X,
which represents a combination of the electron concentration and and
the magnetic field distribution. A search for emission in the EGRET
data base revealed marginal evidence for excess emission from this
direction. De Jager et al. \cite{dea96b} have shown that we should not
expect to see MeV to GeV inverse Compton $\gamma$-rays from the Vela X
SNR if the particles and fields are in equipartition with $B\sim
30\mu$G. This is consistent with the EGRET upper limits, which lie
above the flux expected for equipartition.

%
\begin{figure}
\vspace{2.5truecm}
\centerline{ \hskip -0.5truecm\psfig{file=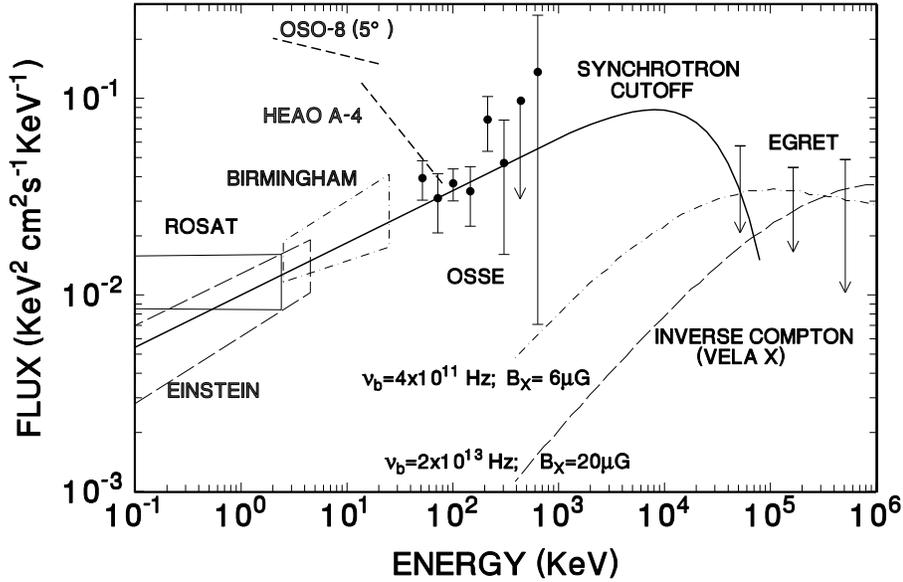,height=6.0in} }
\vspace{-8.6truecm}
\caption{The EINSTEIN, ROSAT, \& Birmingham X-ray spectra of the $1'$ 
Vela compact nebula (see [21] for references).  See also ref. [21] for
the corresponding OSSE spectrum of the Vela $1'$ radius compact
nebula.  The 30-100 MeV EGRET upper limit constrains the extent of this
spectrum to an e-folding cutoff energy $<40$ MeV [20].  EGRET upper
limits above 100 MeV constrain the inverse Compton contribution from
Vela X, with the break frequency $\nu_b$ and field strength $B_X$ as
free parameters. Two such Vela X spectra constrained by EGRET are shown
by the dot-dash and long dashed lines (see ref. [20] for details).  }
\label{fig:Vela}
\vspace{-0.0truecm}
\end{figure}

Fig.\ref{fig:Vela} shows some non-equilibrium inverse Compton spectra
which are just below the EGRET upper limits, and it is clear that we
must know the maximum electron energy (which is seen as a cutoff at a
frequency $\nu_b$ in the radio spectrum) before we can give a
meaningful lower limit on $B$. Unfortunately the detection of $\nu_b$
in the radio to far-infrared region is hampered by the presence of dust
emission in the far IR. Again, a next generation gamma-ray observatory
may lead to a detection of Vela X, which should allow a more detailed
study of the parameters involved.

\subsubsection*{Relic electrons at the birthplace of the Vela pulsar?}
The CANGAROO telescope has reported unpulsed emission above 2.5 TeV
from the Vela pulsar. The peak of this emission is located $8'$
southeast of the pulsar, which is the projected birthplace of the
pulsar, given the 11,000 year spindown age of the pulsar, and the
observed proper motion vector \cite{yos96}. A hint of radio emission at
this birthplace is also seen (Frail, D.A. 1996, private
communication).  Using archival ASCA data, Harding et al.\cite{hdg97}
have shown that the X-ray and TeV $\gamma$-ray observations indicate
the presence of relic electrons left in the trail of the moving pulsar
if the magnetic field strength at this site is between 1.3 and 3
$\mu$G. Such a low field would have allowed electrons to have survived
since the birth of the pulsar.  While the X-ray emission is the result
of synchrotron emission from relic TeV electrons in a weak field, the
TeV $\gamma$-ray emission results from the inverse Compton scattering
of the same electrons on the cosmic microwave background.

\subsection*{The Unpulsed TeV Source: PSR B1706-44.}
Frail, Goss \& Whiteoak \cite{fgw94} identified a $4'$ halo ($\sim 60$
mJy flux density at 1.5 GHz) of extended plerionic radio emission
around the pulsar PSR1706-44. The $20'$ trailing emission, which, if
attributed to proper motion effects, implies that the pulsar and SNR
G343.1-2.3 are not associated. Becker, Brazier, \&
Tr\"umper\cite{bbt95} identified unpulsed power law ($\alpha_x=1.4\pm
0.6$ energy index) X-ray emission from this pulsar, with a spatial
extent which is consistent with the $\sim 0.5'$ spatial resolution of
the ROSAT PSPC.  No soft X-ray emission has been detected from the $4'$
radio halo, and Becker, Brazier, \& Tr\"umper \cite{bbt95} made the
point that the unresolved X-ray source associated with this pulsar is a
compact nebula, similar to the compact nebula of the Vela pulsar.

This pulsar was also seen to be emitting unpulsed TeV $\gamma$-rays
\cite{kifune95}, similar to the Vela TeV detection, which strengthens
the conclusion of the similarity between the Vela and PSR1706-44
compact nebulae. However, no unpulsed $\gamma$-ray emission in the CGRO
range was seen from the direction of PSR1706-44, and the pulsed flux
from the pulsar is consistent with the total flux in the EGRET range
\cite{thom96}.  The non-detection of $\gamma$-rays in the CGRO range is
not surprising: any OSSE type emission is expected to be $\sim 10$
times weaker than the corresponding flux seen from Vela, given the
larger distance to PSR1706-44 and similar source parameters.  The
absence of a radio synchrotron nebula due to relic electrons would also
rule out the possibility of an inverse Compton component in the 100 MeV
to 1 GeV range.

\section*{W44: A composite (shell + plerionic) SNR}

The previous discussion was concerned with the detection of gamma-rays
from pulsar-injected electrons (plerions).  Some remnants are composite,
exhibiting both pulsar/plerionic and shell structures, and we may
expect an interesting interaction of electrons injected by the pulsar
into the shell, where further electron acceleration may take place,
resulting in a bright radio shell.  If this remnant is also interacting
with a molecular cloud, we may expect relativistic bremsstrahlung to
produce gamma-rays in addition to inverse Compton scattering.
Furthermore, a $\gamma$-ray component from cosmic ray proton
acceleration in the shell may also be expected.  In fact, the EGRET
instrument on the Compton Gamma Ray Observatory detected high energy
$\gamma$-rays from the vicinity of radio-bright shell-type supernova
remnants $\gamma$ Cygni, IC443, W44 and W28 \cite{esp96}. These
remnants are all associated with molecular clouds, which provide the
natural target material for the production of $\gamma$-rays via either
relativistic bremsstrahlung, or the spallation products of proton-gas
interactions (discussed below).

\subsection*{Morphological properties of W44}
Fig.~\ref{fig:EGRETsources}b (see next section) gives us an indication
of the morphology of W44: The EGRET $\gamma$-ray source 2EG J1857+0118
\cite{thom96,esp96} is located on the eastern side of this remnant, and
from Fig. 1 of \cite{dm97} it is clear that this remnant is also
interacting with a molecular cloud on the eastern side, as inferred
from $^{13}$CO and other molecular line
observations\cite{woo77,ddc76}.  The presence of $^{13}$CO is
indicative of gas densities in excess of $10^3$ cm$^{-3}$ in localized
interstellar ``baseballs,'' whereas the inter-clump densities are of
the order of $\sim 1$ cm$^{-3}$ or less \cite{rps94}. De Jager \&
Mastichiadis \cite{dm97} found an average hydrogen density of $\sim 50$
cm$^{-3}$ in the shell of the remnant, which is large enough for the
production of $\gamma$-rays via various processes.

Whereas the radio is shell-like, the X-ray emission \cite{rps94} is
centrally-peaked, with the radio pulsar PSR B1853+01 offset from the
center of the remnant. The association of the pulsar with the remnant
is strengthened by the detection of a weak cometary tail (the pulsar
wind nebula) pointing towards the center of the remnant \cite{fra96}.
The transverse speed of the pulsar is comparable to the expansion speed
of the radio shell \cite{fra96}.  Harrus, Hughes, \& Helfand
\cite{hhh96} also discovered a smaller X-ray PWN, but the luminosity of
the PWN is however negligible compared to the total X-ray luminosity of
W44.  Arendt \cite{are89} identified infrared emission from W44,
showing a good spatial correlation with W44. This emission is probably
a result of swept-up dust during the SNR expansion, and the infrared
energy density is larger than the other galactic radiation fields
\cite{dm97}. This radiation field was included in the inverse Compton
calculations of de Jager \& Mastichiadis \cite{dm97}.

\subsection*{Spectral properties of W44}
\begin{figure}
\vspace{1.3truecm}
\centerline{ \hskip -1.0truecm\psfig{file=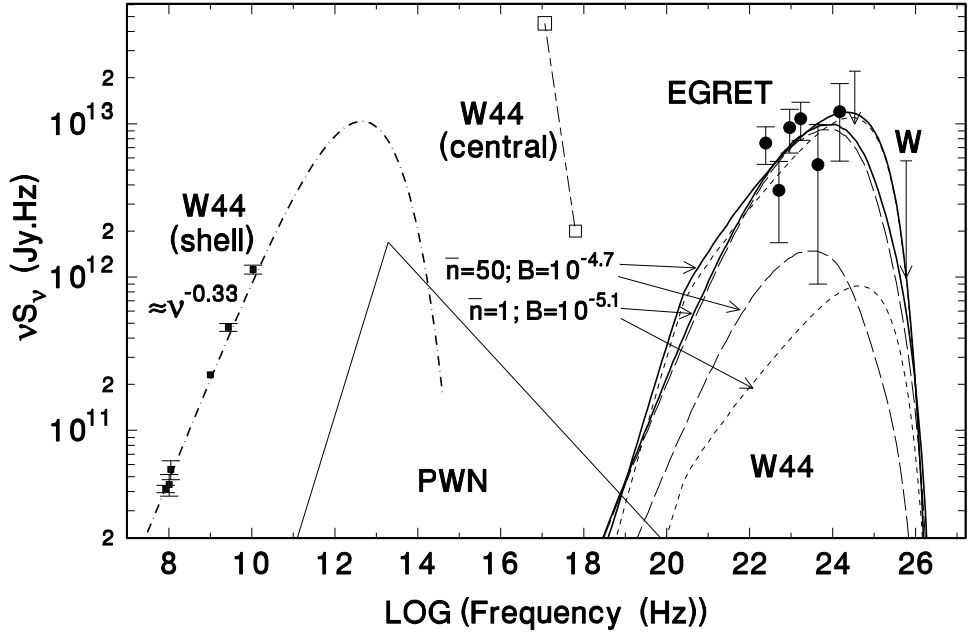,height=5.3in} }
\vspace{-5.9truecm}
\caption{The observed multiwavelength spectrum of W44. See ref. [22]
for references to data/model fits at all wavelengths. The dot-dash line
indicates a model fit to the radio data, which includes a cutoff at
$\nu_b=3\times 10^{12}$ Hz; open boxes --- thermal X-ray energy fluxes
of W44 (the thermal turnover at $\nu<kT/h$ is not included); solid
circles with error bars --- EGRET energy fluxes; upper limit ``W'' ---
Whipple upper limit at $\nu=9\times 10^{25}$ Hz.  {\it Observed
synchrotron spectrum of the pulsar wind nebula}:  solid triangle with
base between $10^{11}$ Hz and $10^{20}$ Hz, with intersection at
$\nu_p=2\times 10^{13}$ Hz. {\it Fits to $\gamma$-ray data} (for two
choices of $\overline{n}$ and $B$ as indicated): short dashed lines ---
relativistic bremsstrahlung; long dashed lines --- inverse Compton;
thick solid lines --- bremsstrahlung + inverse Compton.}
\label{fig:W44}
\vspace{-0.0truecm}
\end{figure}

Fig.~\ref{fig:W44} shows the spectral properties of W44 as summarized
by \cite{dm97}.  The emission from the shell dominates the pulsar
emission (indicated by ``PWN'' in Fig.~\ref{fig:W44}). Whereas the weak
radio/X-ray plerionic component is non-thermal, only the radio
component of the bright shell is non-thermal. The absence of
non-thermal X-ray emission associated with the shell indicates that the
radio spectrum must terminate at frequencies $\ll 10^{16}$ Hz.  The
dot-dash line in Fig.~\ref{fig:W44} is a model fit through the radio
spectrum, and the turnover at $3\times 10^{12}$ Hz (constrained by TeV
$\gamma$-ray observations) assumes an exponential cutoff in the
electron spectrum as discussed by \cite{dm97}.

\vskip 1.6truecm

%

\subsection*{The bremsstrahlung/inverse Compton origin of 2EG J1857+0118}
De Jager \& Mastichiadis \cite{dm97} have shown that the same electrons
that are responsible for the radio emission can also explain the
observed EGRET $\gamma$-ray spectrum between 70 MeV and 10 GeV. {\it
The required cutoff in the high frequency radio spectrum accounts for
the non-detection of W44 in non-thermal X-rays} \cite{rps94} {\it and
TeV $\gamma$-rays} \cite{less95}. This emission is expected to be
concentrated towards the eastern shell where the molecular densities
are highest. The dust emission will also contribute to a weaker inverse
Compton component \cite{dm97}.

The maximum electron energy in the shell, as derived from
multiwavelength observations, is (in terms of the characteristic
synchrotron frequency $\nu_{12}$ corresponding to the cutoff frequency
$\nu_b$)
\begin{equation}
\label{emax}
E_{\rm max}=0.14\, (B_{-5})^{-1/2}\nu_{12}^{1/2}\;{\rm ergs}.
\end{equation}
This maximum energy is orders of magnitude below the maximum obtained
for SN1006 - another shell remnant which shows evidence for electron
acceleration \cite{md96}. The uncertainty in the exact hydrogen density
makes it difficult to infer the magnetic field strength from coupled
synchrotron and bremsstrahlung equations, but values around $10\mu$G are
expected given the approximate hydrogen densities.  De Jager \&
Mastichiadis \cite{dm97} also found that inverse Compton scattering
dominates relativistic bremsstrahlung for molecular densities around 1
particle cm$^{-3}$, whereas bremsstrahlung dominates for
$\overline{n}\sim 50$ cm$^{-3}$, the density expected for this remnant.

\subsection*{The implications of $\gamma$-rays from W44}
By integrating the electron energy spectrum up to the maximum electron
energy, \cite{dm97} obtained a total electron energy content of $E_{\rm
el}=5.8\times 10^{49}B_{-5}^{-1.33}$ ergs, which is about 6 times
larger than the value found for SN1006 by \cite{md96}. The conversion
efficiency ($\eta_{\rm el}=E_{\rm el}/E_{\rm SN}$) of SN explosion
energy $E_{\rm SN}=6.7\times 10^{50}$ ergs to electrons is then
\begin{equation}
\label{etael}
\eta_{\rm el}=0.087 \left[\frac{6.7\times 10^{50}\;{\rm ergs}}
{E_{\rm SN}}\right]B_{-5}^{-1.33},
\end{equation}
which is relatively high for primary electron acceleration in SNR
shocks\cite{er91,mas96}. De Jager \& Mastichiadis \cite{dm97}
addressed this problem by investigating the total output from the
pulsar PSRB1853+01 since the birth of this $\sim 20,000$ year old SNR.
They found that the shell electrons which we are now seeing in radio
and $\gamma$-ray emission, may have been the result of relic electrons
which have been injected into the SNR since birth, provided that the
initial spindown power of the pulsar exceeded the present spindown
power of the Crab pulsar.  Further support for this interpretation
comes from the fact that the radio (and hence electron) spectrum of the
shell of W44 is hard and Crab-like, rather than reminiscent of the softer
shell-type spectra.  Thus, the radio/$\gamma$-ray spectrum of the shell
of W44 may be the result of relic Crab-like electrons injected into the
shell of W44 during the past 20,000 years.

Proton -- gas interactions may result in the production of pions from
which $\gamma$-rays are expected (see below). This process is believed
to account for the $\gamma$-ray emission from shell remnants
interacting with molecular clouds as observed by EGRET. De Jager \&
Mastichiadis \cite{dm97} have however shown that the average gas
density in W44 is not sufficient to account for a dominant proton
contribution to the EGRET $\gamma$-rays, and the contribution from this
hadronic component is probably not more than 20\%. Extrapolation of
this relatively weak component to the TeV range may also account for
the unobservability of W44 at TeV energies, regardless of whether or
not wave damping due to neutrals \cite{ddk96} truncates the proton
spectrum below the TeV range.

\section*{Shell-Type Supernova Remnants}

It has been widely-perceived that supernova remnants (SNRs) are a
principal, if not the predominant, source of galactic cosmic rays (e.g.
see\cite{lc83}) up to energies of around \teq{\sim 10^{15}}eV, where
the so-called {\it knee} in the spectrum marks its deviation from
almost pure power-law behaviour.  Such cosmic rays are presumed to be
produced by diffusive shock (Fermi) acceleration in the environs of
supernova shocks.  Remnants are a convenient origin for cosmic rays
below the knee because their ages (between 100 and \teq{10^5} years)
and sizes permit the diffusive process to accelerate up to such high
energies, they have the necessary power to amply satisfy cosmic ray
energetics requirements, and current estimates of supernova rates in
our galaxy can adequately supply the observed cosmic ray density (e.g.
see \cite{be87}).

The evidence for cosmic ray acceleration in remnants is, of course,
circumstantial.  Nevertheless, the ubiquity of polarized, non-thermal
radio emission in remnants (e.g. see references in the SNR compendium
of Green\cite{dgreen95}) argues convincingly for efficient acceleration
of electrons if the synchrotron mechanism is assumed responsible for
the emission.  X-rays also abound in remnants, and are usually
attributed to thermal emission from shock-heated electrons (because of
the appearance of spectral lines, e.g. see \cite{bsbs96} for Cas A).
The striking spatial coincidence of radio and X-ray images of
shell-type remnants (e.g Tycho\cite{be87} and SN1006; see \cite{kra96}
for a radio/X-ray correlation analysis for Cas A) suggests that the
same mechanism is responsible for emission in both wavebands.  This
contention has recently received a major boost with the discovery of
non-thermal X-ray emission in SN1006\cite{koy95},
which implies\cite{rey96} the presence of non-thermal electrons at
super TeV energies.  In addition, very recent ASCA spectra (Keohane et
al.\cite{keo97}) for the remnant IC 443 and RXTE observations of Cas A
(Allen et al.\cite{all97}) exhibit non-thermal X-ray contributions,
adding to the collection of super-TeV electron-accelerators.  A nice
review of radio and X-ray properties of SNRs is given in\cite{ell94}.

A product of cosmic ray acceleration in SNRs is that such energetic
particles can generate gamma-rays via interactions with the ambient
interstellar medium (ISM), just as in models of the diffuse gamma-ray
background\cite{hun97,hks97}.  Despite theorists' early
expectations\cite{hl75,chev77} that remnants will be gamma-ray bright,
no definitive detection of such emission from a supernova remnant has
been reported.  Prior to the launch of CGRO, associations of remnants
with gamma-ray sources has been limited to two unidentified COS-B
sources\cite{poll85} (for $\gamma$-Cygni and W28), however CGRO has
played an important role in advancing this field.  Our emphasis in the
discussions below will be on shell-type remnants such as
$\gamma$-Cygni, IC 443 and W28.

\subsection*{Gamma-Ray Observations in the CGRO Era}

\begin{figure}
\vspace{0.2truecm}
\centerline{ \hskip 3.90truecm\psfig{file=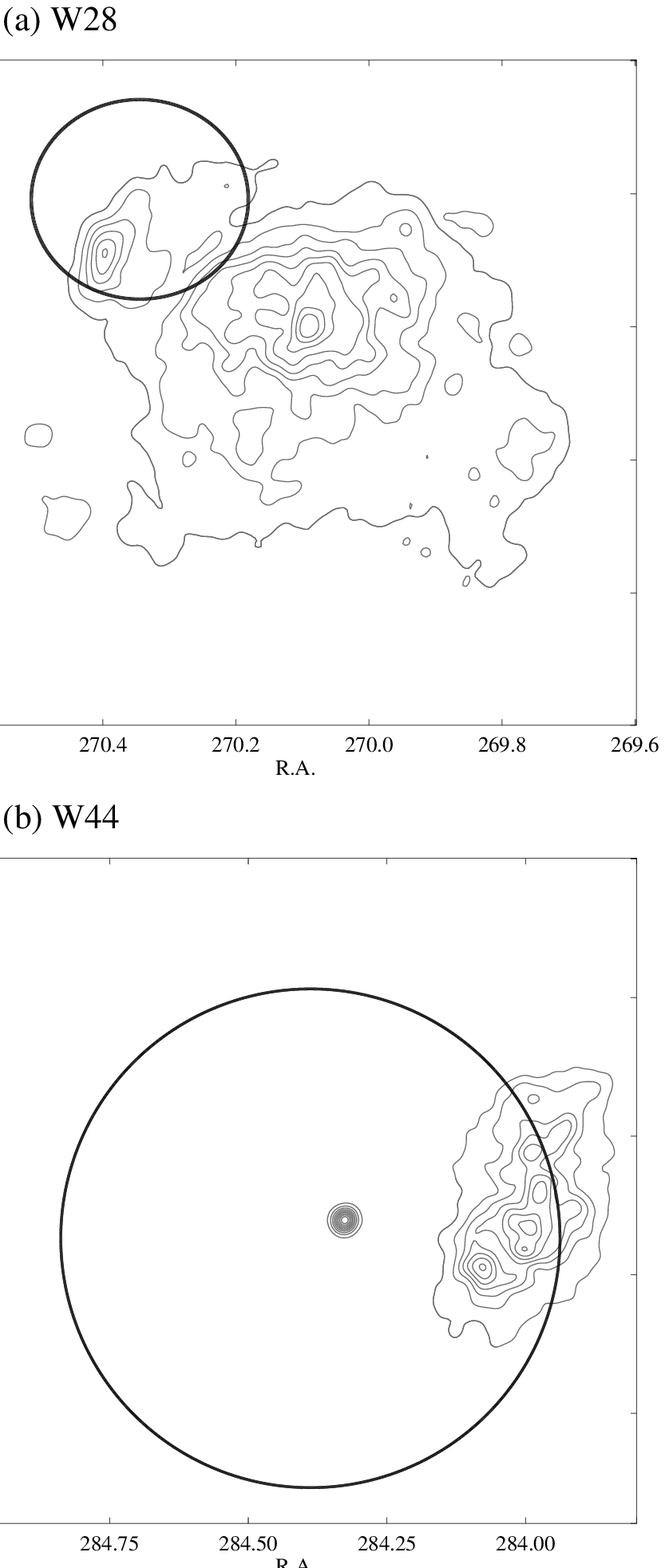,height=5.0in}
              \hskip -2.0truecm\psfig{file=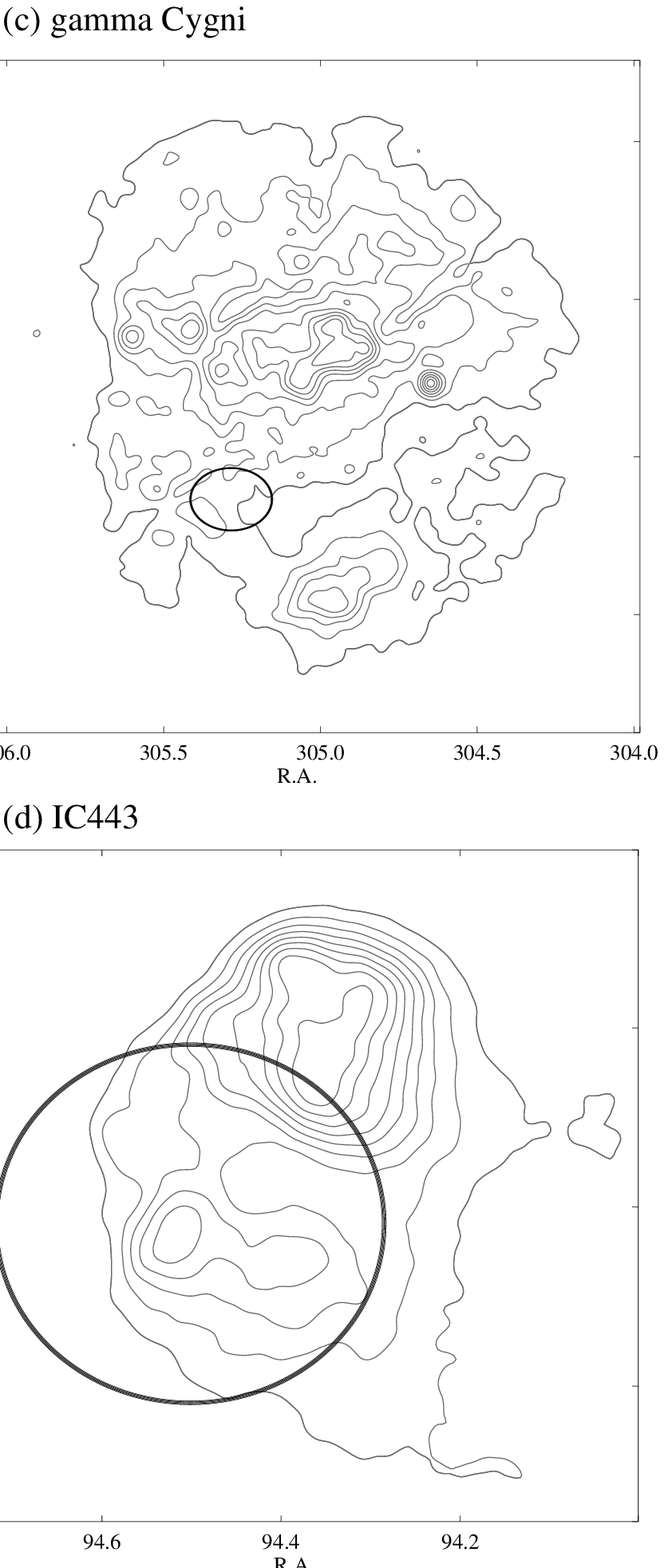,height=5.0in}}
\vspace{0.2truecm}
\caption{
The X-ray/gamma-ray images of the supernova remnants W28, W44,
$\gamma$-Cygni and IC443, as presented in Esposito et al. [34].  
These consist of X-ray contours from the ROSAT
telescope's HRI, and the 95\% (elliptical) confidence contours of
emission above 100 MeV in the associated EGRET unidentified sources.
}
\label{fig:EGRETsources}
\end{figure}

The potential importance of the Compton Observatory's contribution to
the study of supernova remnants was identified on a theoretical level
by Drury, Aharonian and V\"olk\cite{dav94}, and in an observational
context by Sturner and Dermer\cite{sd95} in relation to unidentified
(UID) EGRET sources.  While \cite{sd95} indicated that the latitudinal
distribution of the UID sources may suggest a supernova origin
(discussed more extensively in \cite{mgt97}), it was also pointed out
that the chance probability of coincidental association for a handful
of unidentified EGRET sources with known radio SNR counterparts was
small.  The best candidates for such associations are presented in the
work of Esposito et al.\cite{esp96}, who focused on unidentified EGRET
sources (with approximately \teq{E^{-2}} spectra at above 100 MeV) in
or near the galactic plane and proximate to relatively young
radio/optical/X-ray-emitting remnants.  Such associations, which are at
first glance very enticing, suffer from the large
uncertainty\cite{esp96} in exact directional location of the (assumed
point) sources for the EGRET detections, of the order of 0.5--1
degrees, i.e. the size of typical nearby remnants (see the images
depicted in Figure~\ref{fig:EGRETsources}).  Hence a definitive
connection between {\it any} of these gamma-ray sources and the young
SNRs is not yet possible.

The situation is complicated by the presence of a pulsar (PSR B1853+01)
in the field\cite{esp96,dm97} of the 95\% confidence contour of the
EGRET source 2EG J1857+0118, whose association with the remnant, W44,
is discussed above.  Such a pulsar, or its wind nebula, could easily
generate the observed gamma-ray emission, although no evidence of
pulsation exists in the EGRET data\cite{thom94}.  There is also the
recent suggestion\cite{brkc97} of a pulsar counterpart to the CTA 1
remnant's EGRET source 2EG J0008+7307.  In addition, the improvement of
the localization of 2EG J2020+4026 by the consideration of only
super-GeV photons leads to the conjecture\cite{braz96} that this source
is not associated with the shell of $\gamma$ Cygni, but rather with a
distinct ROSAT source that may also be a pulsar.  The possibility that
pulsars could be responsible for most unidentified EGRET sources near
the galactic plane\cite{kc96} (see also \cite{mgt97}) currently
precludes any assertions stronger than just suggestions of a
remnant/EGRET source connection.  Notwithstanding, it is quite possible
that such remnants could plausibly emit gamma-rays at levels below
EGRET's sensitivity.

%

The absence of TeV emission associated with the remnants surveyed by
Esposito et al. \cite{esp96}, as reported by the Whipple
collaboration\cite{less95,buck97}, is as important for this field as
the detections embodied in the EGRET unidentified sources.  Upper
limits obtained by Whipple to a number of remnants can severely
constrain models, dramatically impacting the hypothesized contributions
of hadronic processes, bremsstrahlung and inverse Compton scattering.
Very recently, the CANGAROO experiment detected\cite{tani97,kif97} the
barrel-shaped remnant SN1006 above 3 TeV, a source for which the
existence of highly super-TeV electrons has already been
established\cite{koy95,rey96} from the X-rays.  Future observations in
the TeV band will provide powerful model diagnostics.

\subsection*{Modelling of Gamma-Ray Production in SNRs}

The modelling of $\gamma$-ray emission from supernova remnants was
limited to a few very preliminary analyses (e.g.
\cite{hl75,chev77,bc82}) prior to the launch of CGRO.  This field of
research began in earnest with the seminal paper of Drury, Aharonian
and V\"olk\cite{dav94}, who computed (as did ref.\cite{nt94}) the
photon spectra expected from the decay of neutral pions generated in
collisions between shock-accelerated ions and cold ions in the ISM.
Since then, there has been a small flurry of activity, with different
groups using alternative approaches, extending the considerations to
include bremsstrahlung and inverse Compton emission.  Here we review
the handful of SNR gamma-ray emission models developed over the last
four years that invoke shock acceleration.

Drury et al.\cite{dav94}, determined the photon spectra and fluxes
expected from the decay \teq{\pi^0\to\gamma\gamma} of neutral pions
generated in hadronic collisions (\teq{pp\to p\pi^0X} etc.) between
power-law shock-accelerated ions and cold ions in the ISM; they
neglected the other (electromagnetic) processes mentioned just above.
Due to the isotropy of decay in the pion rest frame, the decay
kinematics yield\cite{steck71} a photon spectrum that is symmetric
about \teq{m_{\pi}/2\approx 67}MeV, an unmistakable signature of the
production of pions in astrophysical systems (see \cite{hun97,hks97}
for its role in determining the gamma-ray background spectrum).
Supernova remnant models of pion production and decay normally use some
variant of a hybrid approach (e.g. see Dermer\cite{derm86}), where low
energy pion creation (for shock-accelerated proton momenta \teq{p_p}
below around 3 GeV/c) is mediated by an isobaric state \teq{\Delta
(1232)} (Stecker\cite{steck71}), or a collection of different states,
and the complexities of pion creation at high energies (for
\teq{p_p\gtrsim 10} GeV/c) is described by some
adaptation\cite{sb81,tn83} of Feynman scaling.  Drury et al.  used the
{\it two-fluid} approach\cite{dv81,drury83,bar97} to explore shock
acceleration hydrodynamics, treating the cosmic rays and thermal ions
as separate entities (electrons go along for the ride).  This
technique, which is extremely useful for time-dependent problems such
as SNR expansions, obtains solutions that conserve particle number,
momentum and energy fluxes, thereby describing some of the non-linear
effects\cite{je91,bar97} of diffusive shock acceleration.  Drury et
al.  determined that the luminosity peaked in the Sedov phase, in
accord with maximal shock dissipation arising when the supernova ejecta
is being compressed and significantly decelerated by the ISM.

In order to match the EGRET flux, Drury et al.'s pion decay model
requires a {\it high target density} (\teq{>100}cm$^{-3}$).  Gamma-ray
bright remnants might therefore be expected to border or impinge upon
dense regions of the ISM, perhaps giant molecular clouds, in accord
with the earlier suggestions of \cite{bc82}.  Drury et al.
\cite{dav94} predicted that such remnants should become limb-brightened
with age, an effect that arises because, as the shock weakens with
time, the dominant $\gamma$-ray flux is always ``tied'' somewhat to a
region near the shock.  While such a limb-brightening is seen in radio
and X-ray images of remnants (e.g. Tycho and SN1006), higher angular
resolution observations are needed in gamma-ray telescopes before its
existence, or otherwise, can be probed at high energies.  Such a
definitive connection of the $\gamma$-ray emitting regions to a
remnant's shell (which may be ruled out for $\gamma$ Cygni according to
\cite{braz96}) would argue strongly for a shock-acceleration origin of
the energetic particles responsible for emission in the gamma-ray and
other wavebands.  Note that Drury et al. did not incorporate physical
(spatial and temporal) limits
imposed\cite{lc83,md96,rey96,bar97,berg97} on the shock acceleration
mechanism by the supernova shell, so that they permitted particles to
be accelerated to at least 100 TeV.   This omission promoted
observational investigations by the Whipple collaboration that produced
upper limits in the TeV energy range\cite{less95,buck97} that
contradicted the Drury et al. predictions.  While this conflict has
been proposed as a failure for shock acceleration models of SNRs,
realistic choices\cite{dm97,ssdm97,berg97} of the maximum energy
\teq{\Emax} of particle acceleration actually yield model spectra that
are quite compatible with Whipple's observational constraints to
$\gamma$ Cygni and IC 443.

A number of substantial model developments has ensued since Drury et
al.'s enunciative work.  Among these was the work of Gaisser, Protheroe
and Stanev\cite{gps97}.  They computed emission fluxes and luminosities
for the decay of \teq{\pi^0}s produced in hadronic collisions,
bremsstrahlung and inverse Compton scattering, however they omitted
consideration of non-linear shock dynamics in any form, did not treat
time-dependence, and assumed test-particle power-law distributions of
protons and electrons.  In their model, the inverse Compton scattering
used both the microwave background and an infrared/optical background
field local to the SNRs as seed soft photons.  Their bremsstrahlung
component was due to cosmic ray electrons colliding with ISM protons.
Their model has difficulty with the TeV upper limits obtained by
Whipple, unless sufficiently steep particle distributions are
assumed.   Gaisser et al. imposed a high matter density (\teq{>
300}cm$^{-3}$) to enhance the bremsstrahlung and \teq{\pi^0} decay to
inverse Compton flux ratio, thereby generating steeper spectra for the
sources associated with $\gamma$ Cygni and IC443.  Note that for all
models discussed here, the \teq{\pi^{\pm}\to e^{\pm}} secondaries are
always unimportant for the SNR problem since the ion cooling time in
pion production is much longer than typical remnant ages.

Recently Sturner et al.\cite{ssdm97} have developed a time-dependent
model, where they solve for electron and proton distributions subject
to cooling by inverse Compton scattering, bremsstrahlung, \teq{\pi^0}
decay and synchrotron radiation. Like Gaisser et al., the
work of Sturner et al. assumes canonical power-laws but does not
include any treatment of non-linear shock acceleration effects.  One
feature of their model is the dominance of inverse Compton emission,
which intrinsically has a flatter spectrum than either bremsstrahlung
or pion decay radiation.  This arises because they generally opt to
have the same energy density in non-thermal electrons and protons, so
that the shock-accelerated electrons are more populous than their
proton counterparts.  Sturner et al.'s work introduced cutoffs in the
distributions of the accelerated particles (first done by \cite{md96}),
which are defined by the limits on the achievable energies \teq{\Emax}
in Fermi acceleration.  Hence, given suitable model parameters, Sturner
et al. can accommodate the constraints imposed by Whipple's upper
limits\cite{less95} to $\gamma$ Cygni and IC 443.

The most recent development among gamma-ray production models has been
the work of Baring et al.\cite{beg97,berg97} on the application of
non-linear shock acceleration theory to the SNR problem, an appropriate
step given that remnant shocks are strong enough that the generated
cosmic rays are endowed with a significant fraction of the total
particle pressure.  This work utilizes the steady-state Monte Carlo
simulational approach (described in the reviews of \cite{je91,bar97}),
a kinematic technique that can self-consistently model the feedback of
the accelerated particles on the spatial profile of the flow velocity,
which in turn determines the shape of the particle distribution.  In
establishing this feedback, the accelerated population pushes on the
upstream plasma and decelerates it before the discontinuity is reached,
so that an upstream {\it precursor} forms, in which the flow speed is
monotonically decreasing.  At the same time, the cosmic rays press on
the downstream gas, slowing it down too.  The overall effect is one
where the total compression ratio \teq{r}, from far upstream to far
downstream of the discontinuity, actually {\it exceeds that} (i.e. 4)
{\it of the test-particle scenario}, the case where the canonical
power-laws used in \cite{dav94,gps97,ssdm97} are generated.  This
situation results from the need of the flow to increase \teq{r} to
adjust for energy and momentum escape\cite{eich84,ee84}.  If the
particle diffusive scale (i.e. mean free path \teq{\lambda}) is an
increasing function of momentum, as is expected to be the
case\cite{bar97} based on inferences of particle diffusion from the
Earth's bow shock and also in hybrid plasma shock
simulations\cite{gbse93}, then higher energy particles will sample a
stronger shock, yielding upward curvature in the non-thermal cosmic ray
distribution.  This curvature is important for gamma-ray emission
models, since it introduces enhancements\cite{beg97} in the TeV range
by factors of 2--3 relative to the EGRET range; such increases can be
the difference between detection and non-detection by air \v{C}erenkov
experiments like Whipple, CAT, CANGAROO and HEGRA.  The curvature and
the modification of the flow hydrodynamics depend on each other
intimately in a highly non-linear fashion.  Typical distributions of
particles that are accelerated in {\it cosmic ray modified} shocks are
presented in numerous papers\cite{eich84,ee84,je91,ebj96,berg97}.

The self-consistent Monte Carlo approach to shock acceleration in
\cite{berg97} includes neutral pion decay emission, bremsstrahlung and
inverse Compton emission components.  The cessation of acceleration
above \teq{\Emax\sim} 1 TeV - 10 TeV range caused by the spatial and
temporal limitations of the expanding SNR shell yields gamma-ray
spectral cutoffs that are consistent with the Whipple TeV upper
limits\cite{less95,buck97}.  The Monte Carlo approach generates
particle diffusion scales that are always much less than the remnant's
shock radius (as in \cite{dav94}) so that the effects of shock
curvature can be neglected.  This may also render the lack of
time-dependence in the technique a less relevant limitation.  A
prominent feature of the model of \cite{berg97} is the low value of the
electron to proton ratio above 1 GeV, due to a sensible description of
the injection of thermal electrons into the acceleration process.  This
description, which models the way particles diffuse in turbulent plasma
environments, guarantees\cite{bar97} that the electron distribution is
steep enough at low energies so as to render the $e$/$p$ ratio much
less than unity above 1 GeV.  This determination is entirely consistent
with the observation that electrons supply around 2\% of the cosmic ray
population by number\cite{mull95}, and also blends with limits on the
local $e$/$p$ abundance ratio imposed when modelling the galactic
gamma-ray background\cite{hun97}.  This contrasts the situation of
\cite{ssdm97}.  Note that while bremsstrahlung is more efficient than
pion decay emission for given cosmic ray electron and proton energies,
the emergent bremsstrahlung component can be inhibited if the $e$/$p$
ratio is low.  Future measurements of the unidentified sources by more
sensitive experiments in the 1--100 MeV range should constrain the
$e$/$p$ ratio.

While the focus here has been on gamma-rays from shell-type remnants,
much can be learned from studying other wavebands also.  This has been
the approach of Mastichiadis and De Jager\cite{md96}, who have studied
the remnant SN1006.  For SN1006, which has not been seen in gamma-rays,
they used\cite{md96} the recent observations\cite{koy95} of non-thermal
X-rays by ASCA to constrain the energy of electrons and the magnetic
field, interpreting the X-ray flux as being of synchrotron origin.
This contention (see also Reynolds\cite{rey96}) assumes that the steep
X-ray spectrum is part of a rollover in the electron distribution at
energies around 100 TeV.  Using microwave and infrared backgrounds
appropriate to SN1006, \cite{md96} predicted the resulting inverse
Compton component in $\gamma$-rays, and determined that it would always
satisfy the EGRET upper bounds.  However, they concluded that TeV upper
limits from experiments like Whipple could potentially constrain the
ratio \teq{\eta =\lambda/r_g} of the electron mean free path
\teq{\lambda} to its gyroradius \teq{r_g} to values signifying
departure from Bohm diffusion (i.e. \teq{\eta\gg 1}), otherwise the TeV
flux would exceed that of the Crab nebula.  Such a conclusion appears
to be borne out by the very recent announcement by Tanimori et
al.\cite{tani97,kif97} of the detection of SN1006 above 3 TeV by the
CANGAROO experiment, with the flux at these energies probably being due
to an inverse Compton component.  Pinning the X-ray synchrotron
spectrum determines \teq{\Emax^2 B} and also a combination of \teq{B}
and the electron density, where \teq{\Emax} is the maximum accelerated
electron energy.  Through \teq{\Emax}, \teq{\eta} couples to \teq{B} so
that the gamma-ray inverse Compton flux anti-correlates with both
\teq{B} and \teq{\eta =\lambda /r_g}.  This interplay between the
wavebands (see also \cite{ssdm97}) will play an important role in
future model developments for shell-type remnants.

\newpage

\section*{Conclusion and A Look Ahead}

The Compton Gamma-Ray Observatory has propelled the study of supernova
remnants, plerionic and non-plerionic, into the foreground of gamma-ray
astrophysics.  The various aspects of the plerions, the shell-type
remnants, and the W44 composite discussed in this review serve to
underline the diversity of the handful of definitive or candidate
$\gamma$-ray remnants observed by CGRO.  Such a diversity is also
reflected in their morphological properties, their optical/IR spectra,
environmental densities, etc.  It follows that no two sources seem the
same so that they must be considered on a case-by-case basis.  While
the plerions can easily derive their luminosity from the parent pulsar,
perhaps the proximity of the non-plerionic sources to dense molecular
clouds of various sorts provides a strong clue to the reason for their
$\gamma$-ray emission.  It is clear that if some of the EGRET
detections turn out to be of gamma-rays generated in the environs of
remnant shells, then gamma-ray emitters must be a minority of remnants,
perhaps mostly young, given that they cannot produce ions above around
a few TeV in profusion.  Remnants that provide cosmic rays up to the
knee must consequently be a gamma-ray quiet majority.  Alternatively,
if fluxes of shell origin are well below EGRET's flux sensitivity,
then the notion that shell-type remnants are simultaneously gamma-ray
bright and prolific producers of cosmic rays becomes tenable.  It has
therefore become evident that the Whipple upper limits have not
destroyed the hypothesis that shocks in shell-type remnants energize
the particles responsible for the gamma-ray emission, but rather have
provided a powerful tool for constraining our understanding.  Much
remains to be explored in this field, in particular the relationship
between the clouds and the shock parameters, the degree of ionization
of the environment, the precise location of the gamma-ray emission,
differentiation between plerion-driven and shock-powered gamma-ray
sources, and the maximum energies and relative abundances of the
produced cosmic rays.   The next generation of both space-based and
ground-based gamma-ray telescopes, with better angular resolution and
cumulatively-broad spectral range will have a significant impact on
this field, particularly in coordination with X-ray and radio
observations.

\vskip 9pt\noindent {\bf Acknowledgments:}
We thank our collaborators Alice Harding, Apostolis Mastichiadis, Don
Ellison, Steve Reynolds and Isabelle Grenier for many informative
discussions about supernova remnants and shock acceleration theory.  We
also thank Joe Esposito for providing the images used in
Figure~\ref{fig:EGRETsources}.

\end{document}